\documentclass[twoside]{article}
\usepackage{amsfonts,amssymb,amsbsy,textcomp,marvosym,picins,amsmath,caption,threeparttable,amsthm,subfigure}
\usepackage{eurosym,mathrsfs,fancyhdr,CJK,multicol,graphics,indentfirst,color,bm,upgreek,booktabs,graphicx,listings}
\usepackage{multirow}
\def\footnoterule{\kern 1mm \hrule width 10cm \kern 2mm}

\allowdisplaybreaks
\catcode`@=11
\def\title#1{\vspace{3mm}\begin{flushleft}\vglue-.1cm\Large\bf\boldmath\protect\baselineskip=18pt plus.2pt minus.1pt #1
\end{flushleft}\vspace{1mm} }
\def\author#1{\begin{flushleft}\normalsize #1\end{flushleft}\vspace*{-4pt} \vspace{3mm}}
\def\address#1#2{\begin{flushleft}\vglue-.35cm${}^{#1}$\small\it #2\vglue-.35cm\end{flushleft}\vspace{-2mm}\par}

\catcode`@=11
\def\section{\@startsection{section}{1}{\z@}%
 {-3ex \@plus -.3ex \@minus -.2ex}%
 {2.2ex \@plus.2ex}%
{\normalfont\normalsize\protect\baselineskip=14.5pt plus.2pt minus.2pt\bfseries}}
\def\subsection{\@startsection{subsection}{2}{\z@}%
 {-3ex\@plus -.2ex \@minus -.2ex}%
 {2ex \@plus.2ex}%
{\normalfont\normalsize\protect\baselineskip=12.5pt plus.2pt minus.2pt\bfseries}}
\def\subsubsection{\@startsection{subsubsection}{3}{\z@}%
 {-2.2ex\@plus -.21ex \@minus -.2ex}%
 {1.4ex \@plus.2ex}
{\normalfont\normalsize\protect\baselineskip=12pt plus.2pt minus.2pt\sl}}

\begin{document}
\begin{CJK*}{GBK}{song}
\thispagestyle{empty}
\vspace*{-13mm}
\vspace*{2mm}

\title{Usage Scenarios for Byte-Addressable Persistent Memory in High-Performance and Data Intensive Computing}

\author{Mich\`{e}le Weiland$^{1,*}$ and Bernhard Hom\"{o}lle$^{2}$}

\address{1}{EPCC, The University of Edinburgh, The Bayes Centre, 47 Potterrow, Edinburgh EH8 9BT, UK}
\address{2}{System Vertrieb Alexander GmbH, Paderborn, Ahornallee 9, 33106 Paderborn, Germany}

\vspace{2mm}

\noindent E-mail: m.weiland@epcc.ed.ac.uk; bernhard.homoelle@sva.de\\[-1mm]


\let\thefootnote\relax\footnotetext{{}\\[-4mm]
\indent\ This is a post-peer-review, pre-copyedit version of an article published in Journal of Computer Science and Technology. After publication, the final authenticated version is available online at: http://dx.doi.org/10.1007/s11390-020-0776-8\\[.5mm]
\indent\ The NEXTGenIO (Next Generation I/O for the Exascale) project has received funding from the European Union's Horizon 2020 Research and Innovation programme under Grant Agreement no. 671951. \\[.5mm]
\indent\ $^*$Corresponding Author
}

\noindent {\small\bf Abstract} \quad  {\small Byte-addressable persistent memory (B-APM) presents a new opportunity to bridge the performance gap between main memory and storage. In this paper, we present the usage scenarios for this new technology, based on the capabilities of Intel's DCPMM. We outline some of the basic performance characteristics of DCPMM, and explain how it can be configured and used to address the needs of memory and I/O intensive applications in the HPC and data intensive domains. Two decision trees are presented to advise on the configuration options for B-APM; their use is illustrated with two examples. We show that the flexibility of the technology has the potential to be truly disruptive, not only because of the performance improvements it can deliver, but also because it allows systems to cater for wider range of applications on homogeneous hardware.}

\vspace*{3mm}

\noindent{\small\bf Keywords} \quad {\small byte-addressable, data intensive, memory intensive, persistent.}

\vspace*{4mm}

\end{CJK*}
\baselineskip=18pt plus.2pt minus.2pt
\parskip=0pt plus.2pt minus0.2pt
\begin{multicols}{2}

\section{Introduction}
Byte-addressable persistent memory (hereafter referred to as B-APM) bridges the gap between traditional volatile main memory and persistent storage by offering features, in a single device, that have previously been reserved for either the memory or storage domains. The performance characteristics (such as read and write access rates) of B-APM lie between those of main memory and fast storage devices. 

The broad range of features on offer means that the technology offers great flexibility, but it also means that the configuration of the memory can be complex and have considerable implications on performance; as a result, the use cases are wide ranging. In this paper, we explore the different usage scenarios for B-APM and describe how this type of memory might be used to address bottlenecks in high-performance and data intensive applications. The usage scenarios are based on the specific capabilities of the first true B-APM product, called Optane DC Persistent Memory Modules (DCPMM), released by Intel in April 2019~\footnote{Intel  News  Release,  ``Intel  Announces  Broadest  Product Portfolio   for   Moving,   Storing   and   Processing   Data'', http://newsroom.intel.com/news-releases/intel-data-centric-launch. April 8, 2019}. We believe most of the features of DCPMM, and thus the scenarios described in this paper, will be supported by similar B-APM products in the future, as and when they emerge.

\setcounter{figure}{0}
\begin{figure*}[!htb]
\centering
  \subfigure[]{
    \includegraphics[scale=0.9]{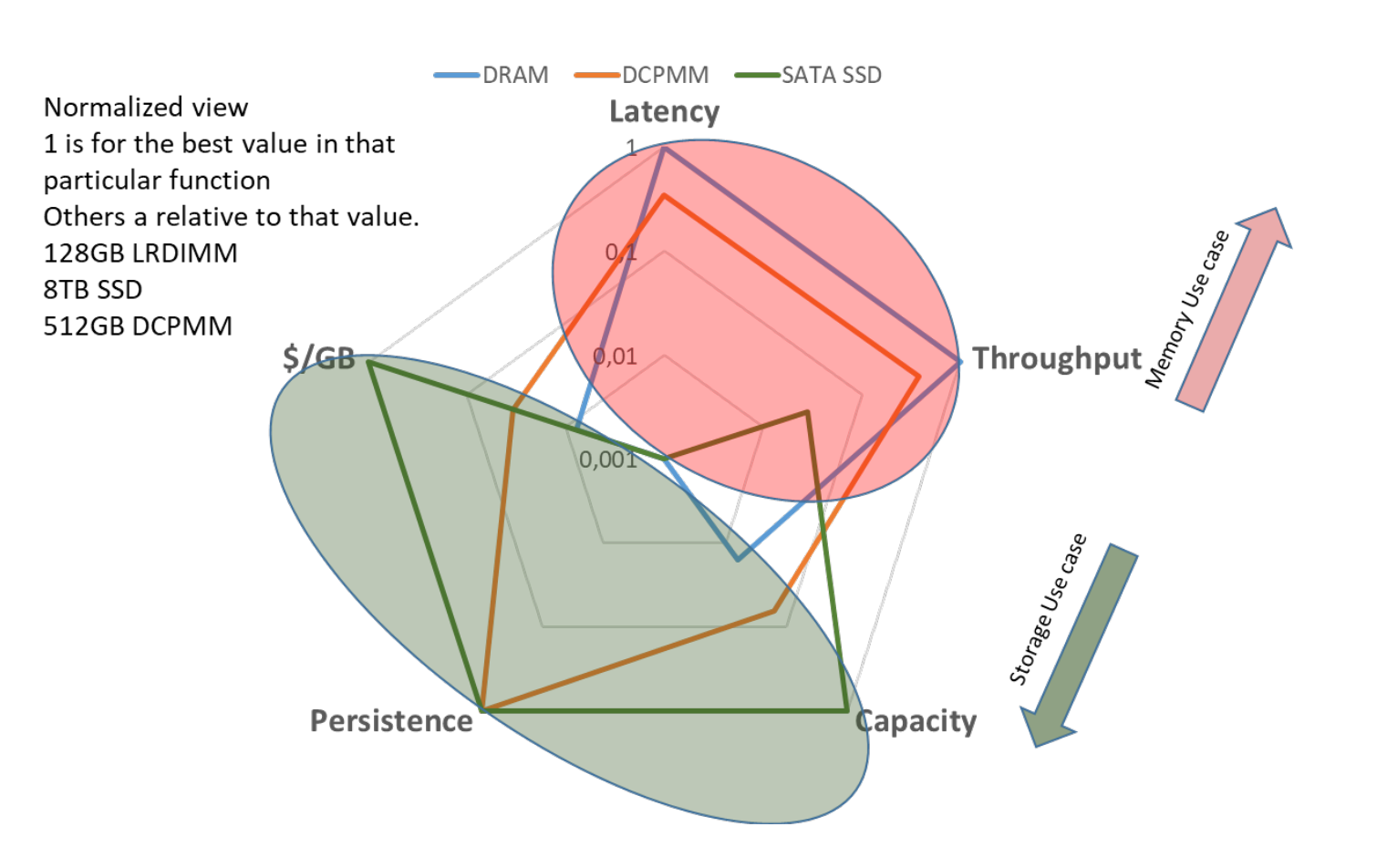}}
  \caption{Graphical representation of the memory and storage spaces in terms of latency, throughput, capacity, persistence and cost, and the position of DRAM, DCPMM and SSDs in those spaces}
  \label{kiviat}
\end{figure*}
\baselineskip=18pt plus.2pt minus.2pt
\parskip=0pt plus.2pt minus0.2pt

\section{Hardware properties}
Since the release of the DCPMM hardware, there have been a number of publications to discuss the hardware properties and features~\cite{jackson}~\cite{izraelevitz}~\cite{patil}~\cite{mason}. However, as it is important for the understanding of this paper, we include a brief overview of the most important characteristics here, together with our own basic performance measurements.

\subsection{Intel Optane DCPMM features} 
Intel Optane DC Persistent Memory Modules (DCPMM) are a type of byte-addressable persistent memory (B-APM) using the DDR4 DIMM form factor. DCPMM is based on the 3D XPoint memory device technology, which implements a cross-point structure that allows each memory cell to be directly addressed by a ``row-column'' selector. The storage medium is transistor-less and leverages 3D stacking to achieve high density and provide much higher capacity than DRAM cells today. The data bits are stored persistently in the media and do not need to be supplied with power to remain set. The main performance features of Intel DCPMM can be summarised as:

$\bullet$ Byte addressable, cache coherent and with load and store accesses without page caching;

$\bullet$ Low latency performance (around ~100 - ~300 ns);

$\bullet$ Large capacity, with up to 512 GB per module;

$\bullet$ Support for Direct and Remote Direct Memory Access (DMA and RDMA);

$\bullet$ Support for storage operations without paging, context switching or interrupts.

In addition, DCPMM offers very high endurance (an expected lifetime of 5 years with 500PB written per module) when compared to DRAM or SSDs.

The Kiviat diagram in {\bf Fig.~1} shows a graphical representation of the memory and storage space in terms of latency, throughput, capacity, persistence and cost, and it compares the coverage of DRAM, DCPMM and SSD for each of these points. While DRAM and SSD only cover the memory and storage aspects respectively, DCPMM is shown to be more generally addressing all points. 

\subsection{Platform Modes of Operation}
B-APM as implemented using DCPMM technology supports two main {\it Platform Modes} of operation: {\it AppDirect} mode and {\it Memory} mode. Switching between the Platform Modes requires a system reboot.

AppDirect mode enables the use of DCPMM as storage. The available storage space can be subdivided by creating namespaces and access to them is similar to memory mapped file operations. In order to support direct access, the file system and the operating system must support the \texttt{dax} mount option, however this is readily available for the standard \texttt{xfs} and \texttt{ext4} file systems. 

Memory mode on the other hand uses the persistent memory as an extension to the main (DRAM) memory. In this mode, the entire main memory is used as the Last Level Cache (LLC) and access to the B-APM is transparent. Importantly, in this mode the persistence of the memory technology is no longer used.

In addition to the Platform Modes themselves, DCPMM can also be partitioned into separate {\it spaces} for Memory mode and AppDirect mode access in order to enable a mixed use of the memory, where part of the B-APM can be used as storage (with different namespaces, if required) and part as memory (see {\bf Fig.~2} and {\bf ~3}). Memory space is however only available in Memory mode. Changing between Platform Modes in a mixed setup will change the environment thus:

$\bullet$ AppDirect mode: the AppDirect space is accessible, but the Memory space is not. Main memory is entirely DRAM.

$\bullet$ Memory mode: the AppDirect space is accessible, and so is the Memory space (transparently). The main memory is the total size of all the Memory spaces in DCPMM; DRAM is not available explicitly, but only as LLC. 

\begin{center}
\includegraphics{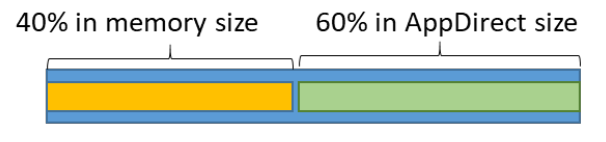}\\
\vspace{2mm}
\parbox[c]{8.3cm}{\footnotesize{Fig.2.~} Illustration of DCPMM divided into AppDirect and Memory spaces}
\end{center}

\begin{center}
\includegraphics{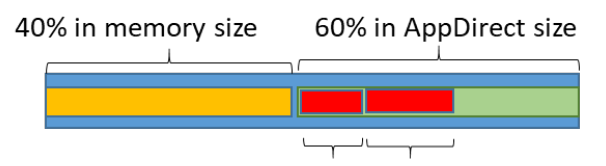}\\
\vspace{2mm}
\parbox[c]{8.3cm}{\footnotesize{Fig.3.~}  Illustration of DCPMM divided into different spaces, with the AppDirect space further split into multiple namespaces }
\end{center}

\subsection{Cache Hierarchy}
In order to understand the performance implications of AppDirect and Memory modes, it is worthwhile looking at the cache hierarchies that each mode exposes. {\bf Fig.~4} is an illustration of the differences in the cache hierarchies, depending on the Platform Mode that is used. 

In {\bf Fig.~4a} the system is in AppDirect platform mode. Data from the main memory as well as from the B-APM will be cached first in the L2 cache before being consumed by the core and corresponding L1 cache. Copy back takes place for any modified as well as unmodified cache lines to the L3 cache. From the L3 cache, only modified cache lines will be copied back into the main memory DRAM or to the DCPMM. 

{\bf Fig.~4b} shows the cache hierarchy for the Memory platform mode. In that case all DRAM is used as cache for the DCPMM. A read from main memory is also a read from DCPMM. This read will fill the L2 cache as well as the Direct mapped DRAM cache. The DRAM cache is a straightforward write back/copy back direct mapped cache. Direct mapped caches are simple, however the replacement policy is strict. With a large DRAM cache of, for example, 192GB, any access in multiple 192GB physical address boundaries will remove all other cache lines. That is true for code as well as data structures that are aligned in such a way.

\setcounter{figure}{3}
\begin{figure*}[!htb]
\centering
  \subfigure[]{
    \includegraphics[scale=0.8]{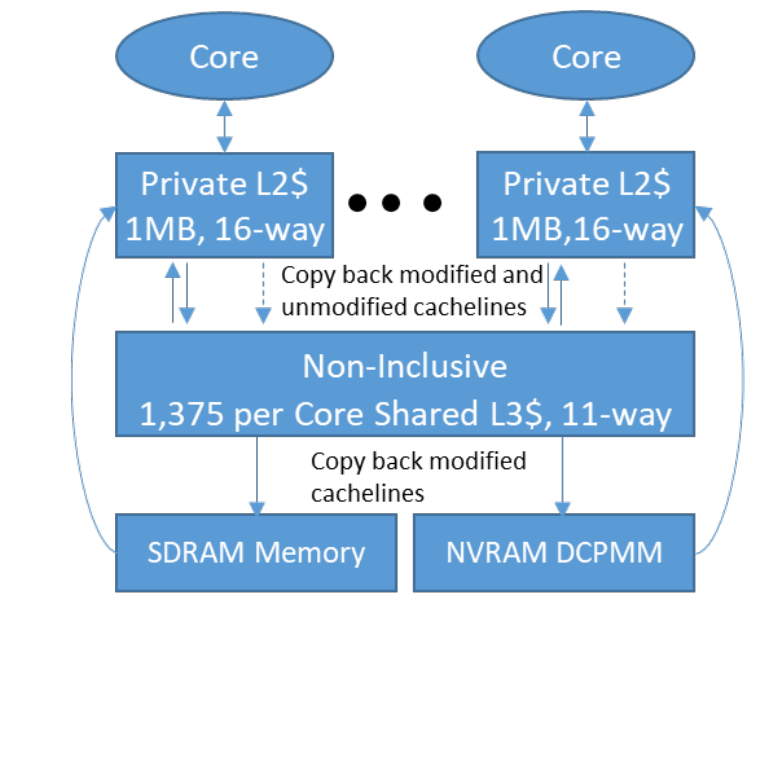}}
  \subfigure[]{
    \includegraphics[scale=0.8]{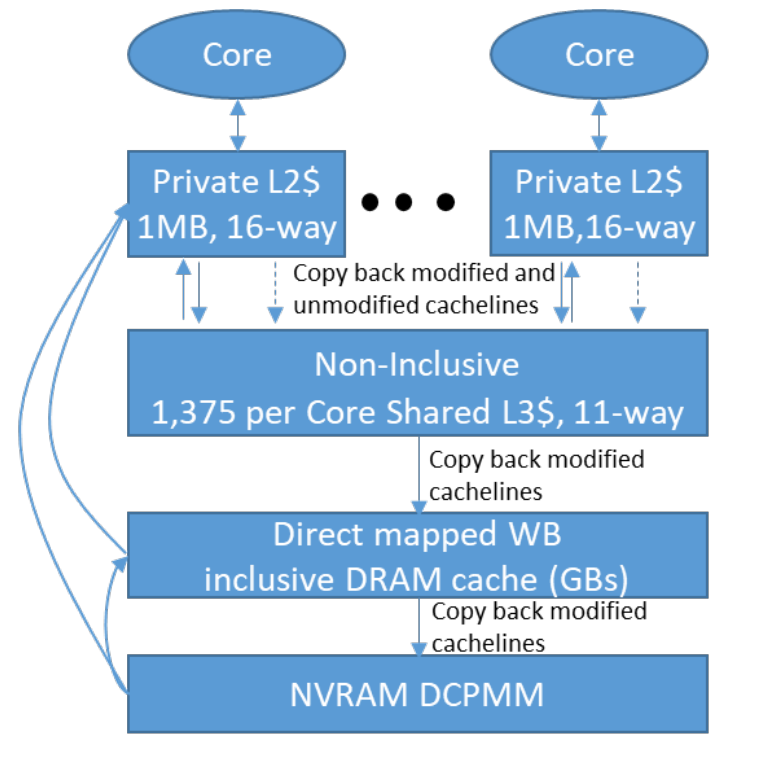}}
  \caption{Illustration of the cache hierarchies for the two Platform Modes. (a) The cache hierarchy for AppDirect mode shows DCPMM sitting next to DRAM. (b) The cache hierarchy for Memory mode shows DRAM becoming the LLC}
\end{figure*}
\baselineskip=18pt plus.2pt minus.2pt
\parskip=0pt plus.2pt minus0.2pt

\subsection{The NEXTGenIO test system}\label{ngiospecs}
The benchmarking system used for all the performance results presented and discussed in this paper is a 34-node prototype system developed as part of the EC Horizon 2020 project NEXTGenIO~\footnote{Next Generation I/O for the Exascale (NEXTGenIO), EC H2020 project, http://www.nextgenio.eu}. Each node has two 24-core Cascade Lake 8260M CPUs, each with six memory channels that can be populated with two DIMMs per channel, resulting in 24 DIMM slots per node. The CPUs are connected through two Ultra-Path-Interconnects (UPI), each with 10.4 GT/s per direction. On each node, half the DIMM slots are filled with 16GB DDR4 DIMMs, and the other half with 256GB DCPMM DIMMs, resulting in a total of 192GB DRAM and 3TB B-APM per node. The system has a small 270TB Lustre file system and uses a custom version of the SLURM resource manager.

The DCPMM DIMMs on each socket are set up using the default ``interleaved'' configuration, which means that the memory banks are used in turn, providing uniform access to the data. The DCPMM are configured to be socket-local, which means that there are NUMA effects to take into account when accessing memory that is {\it not} local to the socket.

\section{Basic Performance}
DCPMM offers much greater capacity that DRAM and is similarly byte-addressable, but the technology does not have the same read or write performance, and this needs to be taken into account when deciding on how to use the memory in different scenarios. This section outlines some of the basic performance characteristics of DCPMM in both AppDirect and Memory modes, and compares them to DRAM.

\subsection{Latency}
Latency, the time elapsed from initiating a request for a byte or word in memory until it is retrieved by the processor, is a fundamental measure of the speed of memory, which is important in HPC and data intensive or streaming applications: the lower the latency, the faster a read operation can complete. {\bf Fig.~5} shows the idle latencies, collected using the Intel Memory Latency Checker~\footnote{Intel Memory Latency Checker, https://software.intel.com/content/www/us/en/develop/articles/intelr-memory-latency-checker.html}, when running DCPMM in Memory and AppDirect mode. Where DRAM is being used as a last level cache in front of the DCPMM memory, we observe that the read latencies are comparable, with around a 10\% performance impact from using DCPMM in this mode. The large DRAM cache means that actual reads from the DCPMM are hidden; the 10\% increase in latency is a good indicator for the additional efforts required to maintain the DRAM cache. AppDirect mode does not have the benefit of caching and the idle latencies for reading from an fsdax namespace are entirely caused by access to DCPMM.

\subsection{Bandwidth}
Memory bandwidth is the rate at which data can be read from or written to memory by a processor. The achievable memory bandwidth is almost certainly less than (and is guaranteed not to exceed) the advertised theoretical bandwidth, and a variety of benchmarks exist to measure sustained memory bandwidth using different access patterns. These are intended to provide insight into the memory bandwidth that a system should sustain on various classes of real applications.
The measurements show in {\bf Fig. 6} were taken with the Intel Power and Thermal Utility 1.1~\footnote{Intel Power and Thermal Utility, https://designintools.intel.com/PTUakaMaximumPowerProgramp/stlgrn28.htm}, which benchmarks the sequential read and write performance to DRAM and DPCMM and reports both the average and maximum achieved values.

The measurements show that the bandwidth of the DCPMM is in the range of the bandwidth of DRAM, though as expected there is a reduction in achieved bandwidth. The bandwidth for the system in Memory mode where DRAM is used as cache shows only a minor decrease, because again (as described for the latency measurements) the large DRAM cache hides access to DCPMM. The bandwidth for DCPMM in AppDirect mode is around 50\% of DRAM for sequential reads and 10\% for sequential writes.

\begin{center}
\includegraphics[scale=0.6]{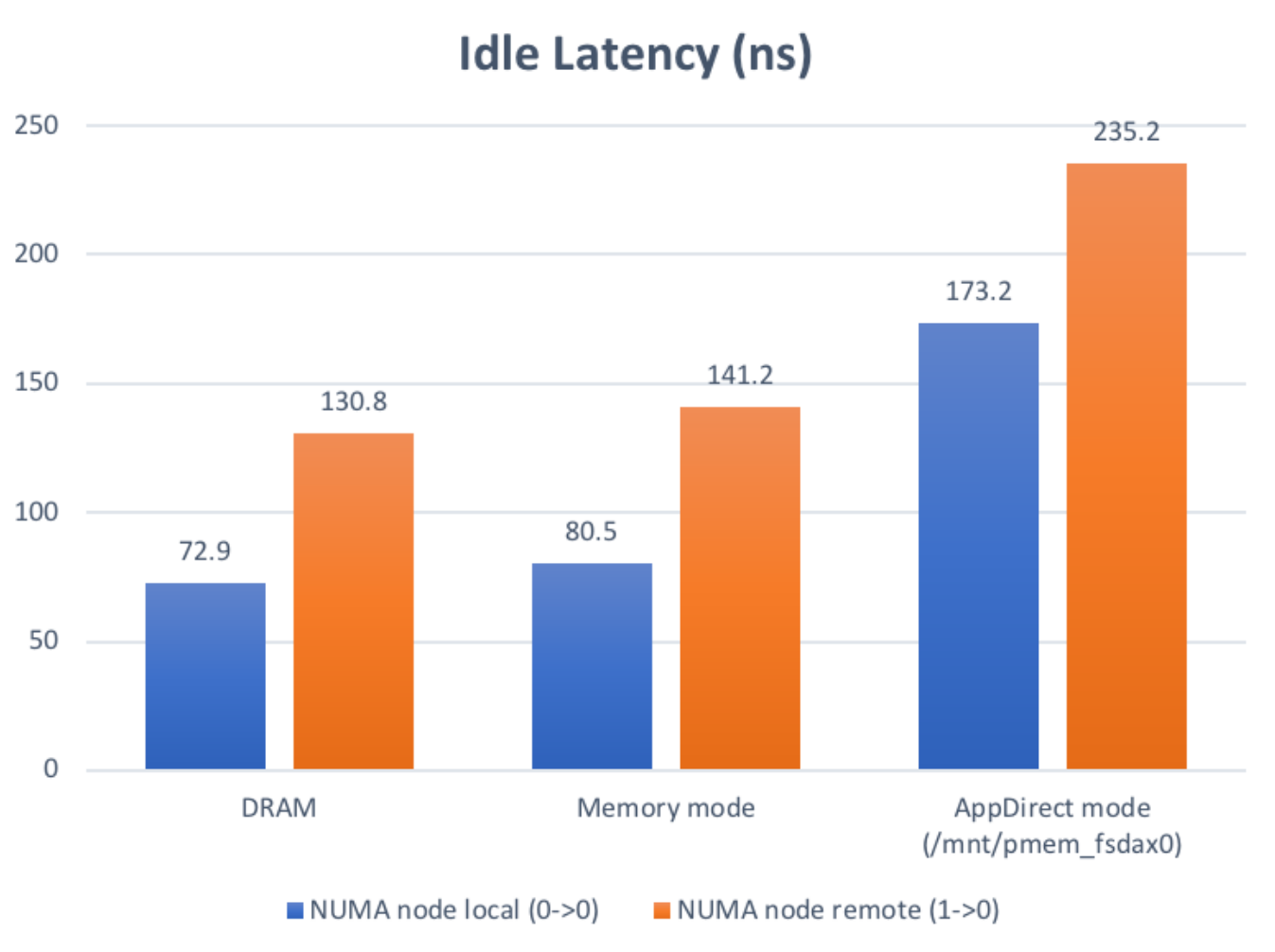}\\
\vspace{2mm}
\parbox[c]{8.3cm}{\footnotesize{Fig.5.~} Idle latency for DRAM and DCPMM in Memory and AppDirect mode. ``Local'' and ``remote'' refers to the NUMA nodes that issue the request, and where the data is requested from}
\end{center}

\begin{center}
\includegraphics[scale=0.6]{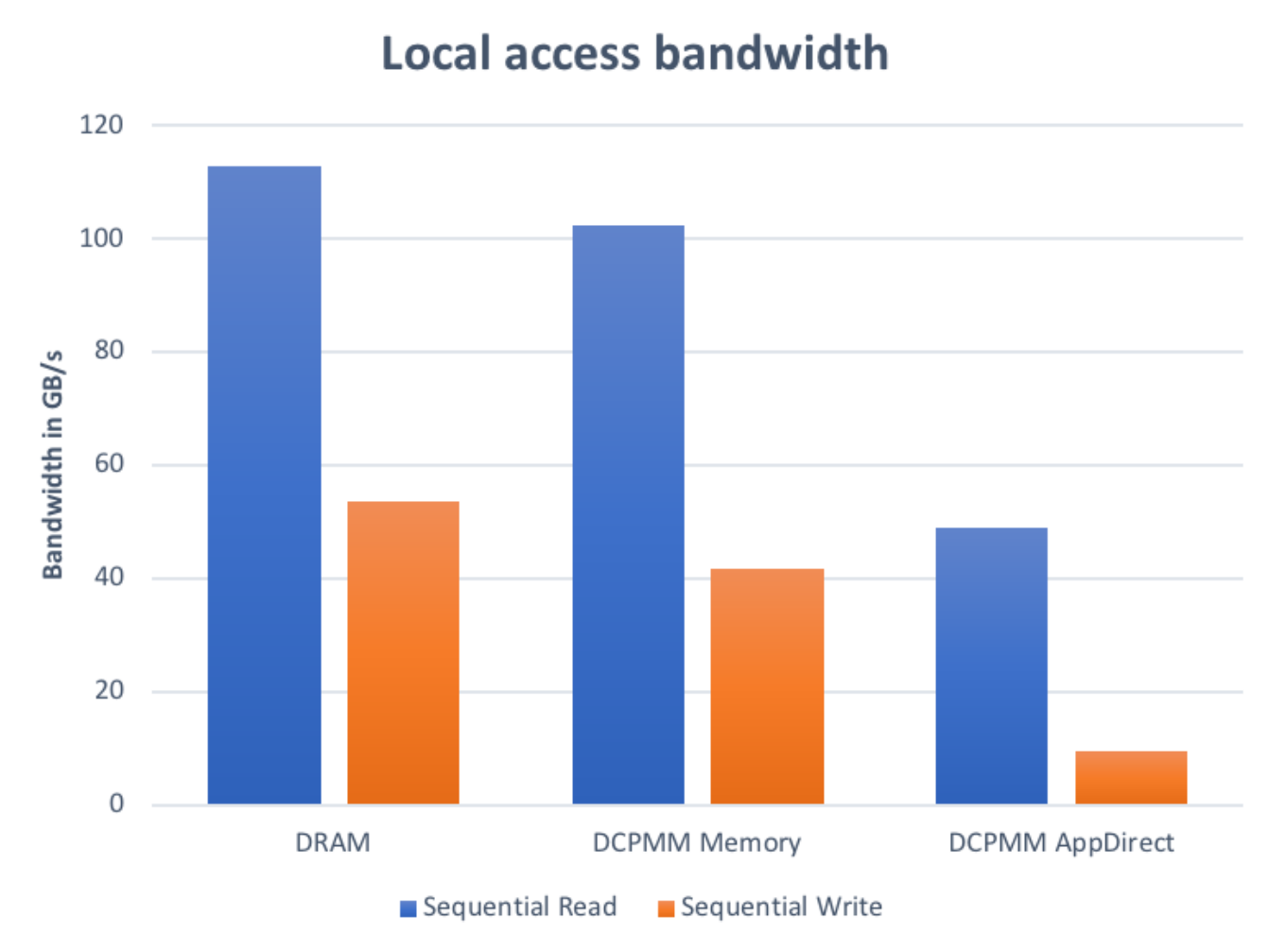}\\
\vspace{2mm}
\parbox[c]{8.3cm}{\footnotesize{Fig.6.~}Measured read and write memory bandwidths for DRAM, DCPMM in Memory mode and DCPMM in AppDirect mode}
\end{center}

\section{Application Usage Scenarios for B-APM}
As shown in the previous sections, B-APM as implemented by DCPMM has a range of features, with varying performance characteristics. In this section, we are going to explore how these features can be used to accelerate applications, taking into consideration not only performance, but also ease of implementation. 

\subsection{Usage Scenarios for Memory Mode}
The easiest usage scenario for B-APM is Memory mode: it is entirely transparent to the user, and no application changes are required at all (not even a recompilation). Memory mode is useful for applications that have a large memory footprint, because the main memory available to the operating system is increased by the amount of Memory space that has been made available. This could be anything from a small portion of the DCPMM capacity to the entirety of it (up to 3TB on the NEXTGenIO test system). A typical way to deal with applications that run out of memory is to either move them to a platform with more DRAM, or to enable them to run over multiple nodes and increase the amount of memory available that way. The former solution is ideal from a performance point of view, however increasing  the amount of DRAM might not be possible due to the cost. The latter solution is more difficult to achieve, especially for applications that are not yet parallelised, and unless an application can achieve good parallel efficiency, scaling it to large numbers of nodes simply to gain access to more memory is a poor use of resources.

An example of such an application is the CASTEP~\cite{castep} computational chemistry application - for some test cases, such as the large DNA benchmark~\footnote{DNA benchmark for CASTEP, http://www.castep.org/CASTEP/DNA}, this application requires much more memory per compute core than is typically available on an HPC system. As a result, these cases are often run both underpopulated (i.e. not using all the cores per compute node) and on tens or hundreds of nodes. The benchmarking and profiling results for this problem that underline this behaviour are described in detail in~\cite{weilandSC} and are not reproduced here. However, it is worth stating that, in addition to the high memory demands, CPU utilisation is low and the parallel efficiency of such jobs is around 25-30\%, because fundamentally the application has a global all-to-all communication pattern that prohibits good scaling; it is therefore ideally run on as {\it few} compute nodes as possible to keep efficiency high. Having said that, we know from the basic performance measurements that there is a penalty for using DCPMM rather than DRAM: the more memory is needed by the application, and the less the data can be reused from cache, the higher the penalty. {\bf Fig.~7} shows a STREAM benchmark measurement using 48 MPI processes (i.e. one node) for gradually increasing array sizes, starting with an aggregate memory requirement of 1.8GB and doubling this size all the way to 915.5GB. The bandwidth numbers shown are the maximum achieved over 20 repetitions. As expected, a clear drop-off can be observed when the data uses a larger portion of the B-APM and no longer fits into the DRAM cache. STREAM is designed not to reuse cached data, and this measurement therefore represents the worst case scenario, however when using DCPMM as extended main memory it is important to understand this performance implication of DRAM caching. Although using DCPMM as main memory might mean that an application can now be run on a single node rather than multiple nodes, it might be beneficial to increase the number of nodes that are used to enable a larger percentage of data to fit into DRAM cache. In order to be beneficial overall, the performance improvement gained from increasing the DRAM cache usage must outweigh the performance penalty from scaling to multiple nodes.

\setcounter{figure}{6}
\begin{figure*}[!htb]
\centering
\includegraphics[scale=0.5]{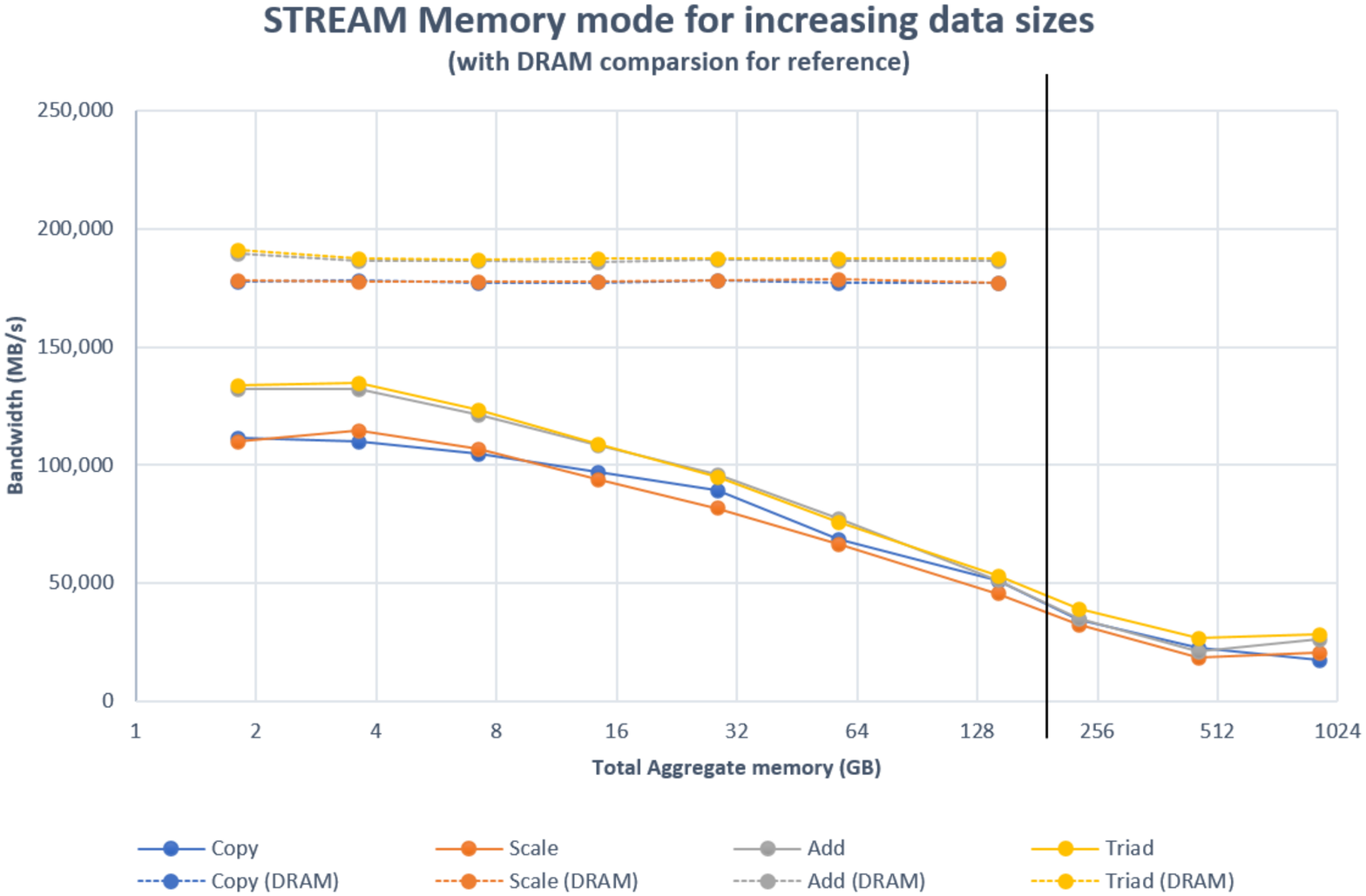}
\caption{STREAM in Memory mode for increasing array sizes, showing a drop in the achieved memory bandwidth that increases with the size of the data structures. This drop levels out once the data no longer fits into the DRAM cache. The equivalent DRAM-only results (dashed lines) are given for reference, and the black vertical line indicates 192GB, the size of the DRAM cache. Note that the x axis is $log 2$}
\label{stream}
\end{figure*}
\baselineskip=18pt plus.2pt minus.2pt
\parskip=0pt plus.2pt minus0.2pt

\subsection{Usage Scenarios for AppDirect Mode}
There are two primary options for using AppDirect mode: firstly, using the namespaces as very fast storage devices similar to SSDs; and secondly, using the namespaces to enable direct access from the application to the B-APM. The first option means that, as for any I/O operation, buffers and page caches will be used. However the second option turns I/O operations into (much faster) direct memory operations, without the use of buffers or page caches. 

\subsubsection{B-APM as local storage}\label{sec:locstor}
The Filesystem-DAX, or {\it fsdax}, mode is the default namespace for DCPMM in AppDirect mode. Simple block devices are created, directly mapped to the B-APM devices, and then mounted as file systems. On the NEXTGenIO test system, the default setup is for two devices per compute node: \texttt{pmem\_fsdax0} is associated with socket 0, and \texttt{pmem\_fsdax1} is associated with socket 1. Applications can write directly to these block devices, with minimal or no changes required to the application. The simplest scenario is that where an application only performs local I/O, i.e. a process only reads and writes data locally without sharing the data with other processes. An example of such an application is the solver step (in this case simpleFOAM) of the OpenFOAM Computational Fluid Dynamics package~\footnote{OpenFOAM, the open source CFD toolbox. https://www.openfoam.com}: its default write strategy is to perform uncollated I/O, which means it creates one directory per parallel process, and all simulation data relevant to that process is written into the process directory. The results shown here were produced using OpenFOAM v1912. The I/O part of an OpenFOAM simulation can be a significant bottleneck because it is meta-data intensive (in addition to creating a directory per process, one directory is created per timestep within each process directory) and predominantly writes small files within the timestep directory. The number of directories and files created scales in line with the number of processes used in the simulation and the timestep write intervals. For example, if each process were to write 8 data files at each timestep during a simulation running on 512 processes for 100 timesteps, outputting data at every step, a total of $500 * 100 * 8 = 400,000$ files would be created and written. 

However, although the simpleFOAM solver writes a lot of simulation data, this data is by default written entirely locally and it is {\it not} read by other processes during the course of the simulation. This is an ideal use case for \texttt{fsdax}: in order to exploit the superior performance of DCPMM, all that is required is to copy the input data to the devices and for each process to write to that same location. It is important to note that the \texttt{fsdax} devices are node local, and as such the input data must be copied to {\it all} the nodes that will be involved in the computation. In addition, on the NEXTGenIO test system each node has two socket-local devices, therefore for optimal performance and to avoid NUMA effects processes should only access the data that is local to them. It is possible to concatenate two devices by linearly mapping them to a single larger device (simply called \texttt{pmem\_fsdax}), however this means losing direct control over data locality.

In order to ensure that processes have access to their local data and only write to their local devices, an easy solution is to use an MDMP style job submission - although verbose, it ensures complete data locality without requiring any changes to the application. A 1-node job (with 24 cores per socket) would be submitted thus using SLURM:

\begin{lstlisting}[
    language=bash,
]
# Copy input data to fsdax0
cd /mnt/pmem_fsdax0
srun -n 1 -n 1 cp -fR /path/to/data .
# Copy input data to fsdax1
cd /mnt/pmem_fsdax1
srun -n 1 -n 1 cp -fR /path/to/data .

# Execute job using MPMD
srun -ppn 48 -n 24 ./exe : -n 24 ./exe
\end{lstlisting}

{\bf Tab.~1} shows the impact on the simpleFOAM solver time per iteration of using either Lustre or fsdax as the write target for a simulation that writes approximately 40MB of data per process at each write interval; the simulation was run for 500 iterations in total. Two sets of experiments are shown: the first shows a setup that increases the write interval from once every 100 time steps to every 1 time step for a simulation that uses 4 nodes. The second uses the same experiment setup, but runs five 4-node simulations at the same time in order to show the effect of I/O contention. For both sets of experiments, and in order to be able to attribute any system performance to the simulations under test, the system was otherwise entirely empty.  

\setcounter{table}{0}
\tabcolsep 9pt
\renewcommand\arraystretch{1.3}
\begin{table*}[!htb]
\centering
\caption{\label{3} Comparing the time (in seconds) per simulation timestep for an OpenFOAM solver with different output intervals, increasing from once every 100 steps to every step, and writing approximately 40MB per process each time (7.5GB per timestep \emph{per simulation}). The write targets are either Lustre or fsdax. The timings are given both for a single 4-node simulation running on an empty system, and for five simultaneous 4-node simulations, using a total of 20 nodes. The data volumes given represent the total amount of data written}\vspace{-2mm}
{\footnotesize
\begin{tabular*}{\linewidth}{l|ccc|ccc}
\toprule
  \multirow{2}{*}{Write interval} & \multicolumn{3}{c|}{single job on 4 nodes} & \multicolumn{3}{c}{ensemble on 5 x 4 nodes} \\
 & Lustre & fsdax & Data volume written & Lustre & fsdax & Data volume written\\ 
 \midrule
 every 100 timesteps (5 writes) & 7.68s & 7.70s & 37.5 GB & 7.67s & 7.70s & 187.5 GB\\ \hline
 every 10 timesteps (50 writes) & 7.89s & 7.87s & 375 GB & 7.82s & 7.84s & 1,875 GB\\\hline
 every 1 timestep (500 writes) & 15.06s & 9.50s & 3,750 GB & 58.97s & 9.32s & 18,750 GB\\
 \bottomrule
\end{tabular*}
}
\end{table*}

It can be seen that, for write intervals 100 and 10 (i.e. writing output 5 and 50 times respectively), the performance for both targets is the same. However, when the I/O volume is increased by another order of magnitude (i.e. writing out 500 times, at every single time step), the effect of the fast node-local storage becomes clear: the small Lustre file system on the NEXTGenIO test bed becomes overwhelmed, and performance (in this case the runtime per iteration) increases by approximately 7x for the ensemble test. Note that this is not to say Lustre as a technology is performing badly per se - at the highest write frequency in the ensemble setup, this experiment writes 18.5TB of data in 3.84 million files and the setup was deliberately chosen as a stress test. Instead this demonstrates that B-APM can be used to easily, i.e. without code changes, absorb a large I/O rate with minimal impact on overall performance when a traditional parallel file system might struggle. For this particular case, the achieved write bandwidth for fsdax is just under 10GB/s.

\subsubsection{B-APM as a distributed file system}
Not all applications have an I/O pattern that lends itself to using the fsdax block devices in the way described above. For example, if multiple processes that reside on different sockets or nodes need to collectively write to the same file, this is a scenario that is not supported as-is. Instead, if such an application wants to be able to use the B-APM without the need for any changes to the application itself, the AppDirect space needs to be presented like a uniform storage layer, much like a network attached file system. This requires either an application-specific data management implementation, or the addition of a system software layer that manages the data and presents it to the application.

One such software layer is GekkoFS~\cite{vef}~\cite{brinkmann}, an ephemeral file system that is created just-in-time for individual jobs, and that sits on top of the B-APM working across nodes. GekkoFS is ephemeral because it is only live for the duration of a job, and it therefore also needs the ability to move data into, and out of, B-APM.

The concept is straightforward: if an application needs access to a distributed storage space, a user can request for a GekkoFS filesystem to be created as part of the job setup and for data to be copied to (and from) it. During runtime, the application will use the GekkoFS filesystem transparently, and no changes need to be made to the application. At the end of the simulation, any data that the user has requested to be kept will be copied back to Lustre and the GekkoFS filesystem will be torn down. The performance of the file system can be tuned depending on the I/O access patterns of an application, e.g. depending on whether data is distributed round-robin across the nodes that make up the files system, and in what chunk sizes, however this level of performance tuning is part of the low-level file system implementation and therefore not exposed at user level. 

The IOR~\footnote{IOR benchmark documentation, https://ior.readthedocs.io} ``hard'' test, which forms part of the benchmarks required for the IO500 list~\footnote{The IO500 list, https://www.vi4io.org/io500/start}, is an example of an application that needs access to a single storage device: it performs random reads and writes to a single shared file. The NEXTGenIO system entered the IO500 competition in July 2020 using the DCPMM as a storage layer with GekkoFS. It achieved 6th position for the 10 node challenge, which also represents the top rank for a non-commercial file system, demonstrating that a distributed file system that sits on top of B-APM is a viable solution that does not require an application to be modified. Full details on how the benchmarks were run, as well as data files with the results, are published as part of the final list~\footnote{The 10 Node Challenge list and data, https://www.vi4io.org/io500/list/20-07/10node}.

\setcounter{table}{1}
\tabcolsep 12pt
\renewcommand\arraystretch{1.3}
\begin{center}
{\footnotesize{\bf Table 2.} Published GekkoFS IOR performance results on the NEXTGenIO system for the July 2020 IO500 10-node challenge}\\
\vspace{2mm}
\footnotesize{
\begin{tabular*}{\linewidth}{c|c}
IOR test & maximum measured bandwidth \\\hline\hline\hline
easy read & 71.968 GiB/s \\
easy write & 63.305 GiB/s \\
hard read & 25.546 GiB/s \\
hard write & 3.310 GiB/s \\
\end{tabular*}
}
\end{center}

\subsubsection{B-APM as direct access storage}
Although a system software layer such as GekkoFS will provide an application with everything it needs in terms of functionality, the performance overheads of distributing and serving the data using a generic approach may be prohibitive for some applications. If performance is a key factor and changes to an application are possible, implementing direct access to B-APM directly using functionality provided by the Persistent Memory Development Kit (PMDK)~\footnote{Persistent Memory Development Kit, https://pmem.io/pmdk} will give better performance results as data can be managed specifically according to the needs of the application and with minimal overheads.

ECMWF, the European Centre for Medium-Range Weather Forecasts, have taken this approach in their implementation of the Fields Database (FDB5)~\cite{fdb2017}, a domain-specific distributed object store that support multiple storage back ends. For operational weather forecasting, which operates under very strict time constraints, being able to achieve optimal performance is key. I/O is a significant bottleneck in the forecasting workflow, both in terms of writing the model output and reading it back for post-processing. In its most recent version, FDB5 was extended to support a B-APM backend, together with a remote distributed front-end, in an effort to reduce the impact that I/O has on the throughput of the forecasting model. This work is described in detail in~\cite{fdb2017} and ~\cite{fdb2019}. As the software is used on a daily basis in production and performance is key, modifying the FDB to use the B-APM directly was a worthwhile investment of time and effort. ECMWF implemented a new backend using PMDK, and the DCPMM DIMMs are unified under a single object store using the frontend to handle the dispatching of data requests between nodes. Implementing the object store as a bespoke solution for the particular data sizes and access patterns means the performance can be fine-tuned for the application. The performance of this solution was measured by assessing the data throughput for the FDB5 using the \texttt{fdb-hammer} benchmark utility to generate and write data, and the \texttt{fdb-read} tool to consume that data. The experiment was carried out using varying numbers of write and server processes. Using 16 server nodes, or 32 server processes, on the NEXTGenIO test system, sustained write performance of 72GiB/s was demonstrated. More details on the experiment setup and results can be found in~\footnote{ H.C. Hoppe et al., ``Final NEXTGenIO Technology Assessment'', September 2019. https://cordis.europa.eu/project/id/671591/results} and~\footnote{T. Quintino et al., ``Running ECMWF's Workflow on the NEXTGenIO Prototype'', September 2019. https://events.ecmwf.int/event/143/contributions/940}. Unlike a B-APM filesystem like GekkoFS, which must be able to handle the distribution of data across the nodes without any specific knowledge of an application's access pattern to that data, a direct access implementation such as FDB5 can take into account such \emph{a priori} knowledge and optimise for best possible read and write throughput. The two methods present a direct trade-off between ease-of-use convenience, and optimal performance.

\subsection{Usage Scenarios for Mixed Setup}
It is possible to configure B-APM nodes to have a mixed setup, with both Memory and AppDirect spaces being available at the same time. This is a very attractive setup for general purpose HPC systems, because it can support applications that are memory intensive, I/O intensive, or both - all using the same hardware. On a system like the NEXTGenIO test bed, splitting 3TB of B-APM into 1TB Memory space and 2TB AppDirect space means that the available main memory can be increased by a factor of 5 (from 192GB to 1TB), while still retaining 2TB of node local storage space, thus satisfying the demands for a very wide range of applications. Changing nodes to different setups (e.g. to AppDirect only) simply requires them to be rebooted with a different configuration, which can be supported through the resource manager.

\subsubsection{Workflows with data dependencies}
The previous sections considered single application scenarios only, however increasingly simulations on HPC systems are moving away from that model and towards a multi-step workflow approach that incorporates data analytics as well as pre- and post-processing~\cite{weilandJSFI}. On traditional HPC systems, one of the key bottlenecks of data intensive workflows that have (data) dependencies between workflow steps is that data has to be copied to and from the network attached file systems between steps. The advantage of node local storage such as B-APM is that it is possible to leave data on the compute nodes. The NORNS~\cite{norns} system software layer, which is integrated with the SLURM resource manager, supports the asynchronous staging of data, and data marshalling between workflow components. As part of the job submission, a user can specify the data dependencies between producer-consumer style workflow components. Based on this specification of what data will be written or read, and whether the data must be stored long term or can be discarded at the end of the workflow, the scheduler ensures that the data is moved to the correct storage layer or node in time for jobs to time. There is no technical reason that prevents this scenario from being implemented using node-local SSDs or any other type of storage, however B-APM enables mixed workflows of memory and I/O intensive applications on the same hardware for the first time. {\bf Fig.~8} illustrates what such a workflow might look like when combining different platform modes and I/O patterns in a single chain of applications with data dependencies between each step.

\begin{center}
\includegraphics[scale=0.5]{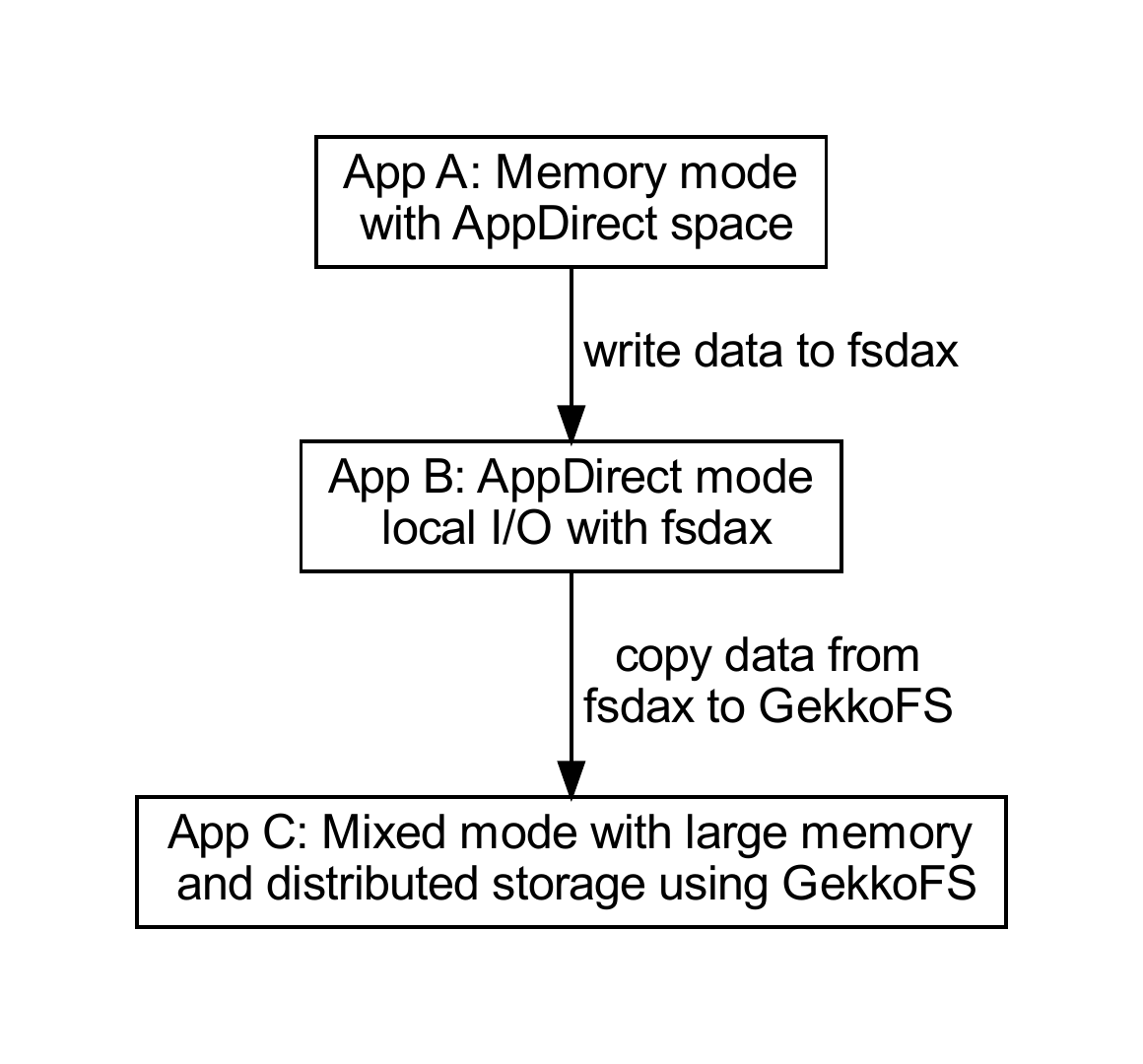}\\
\parbox[c]{8.3cm}{\footnotesize{Fig.8.~}Illustration of the type of workflow that can be supported by a system with B-APM, serving both memory and I/O intensive applications with data dependencies between them}
\end{center}

\subsection{Summary of usage scenarios}
As can be seen from the previous sections, choosing a suitable configuration of B-APM for an application depends on its performance requirements and constraints, its demands on the memory and I/O subsystems, whether or not it can be modified, and what the optimisation goals are. {\bf Fig.~9} shows a simple decision tree guide that summarises the different platform mode choices that are available. {\bf Fig.~10} summarises some of the choices that are available for using B-APM as a fast storage layer.

\setcounter{figure}{8}
\begin{figure*}[!htb]
\centering
\includegraphics[scale=0.5]{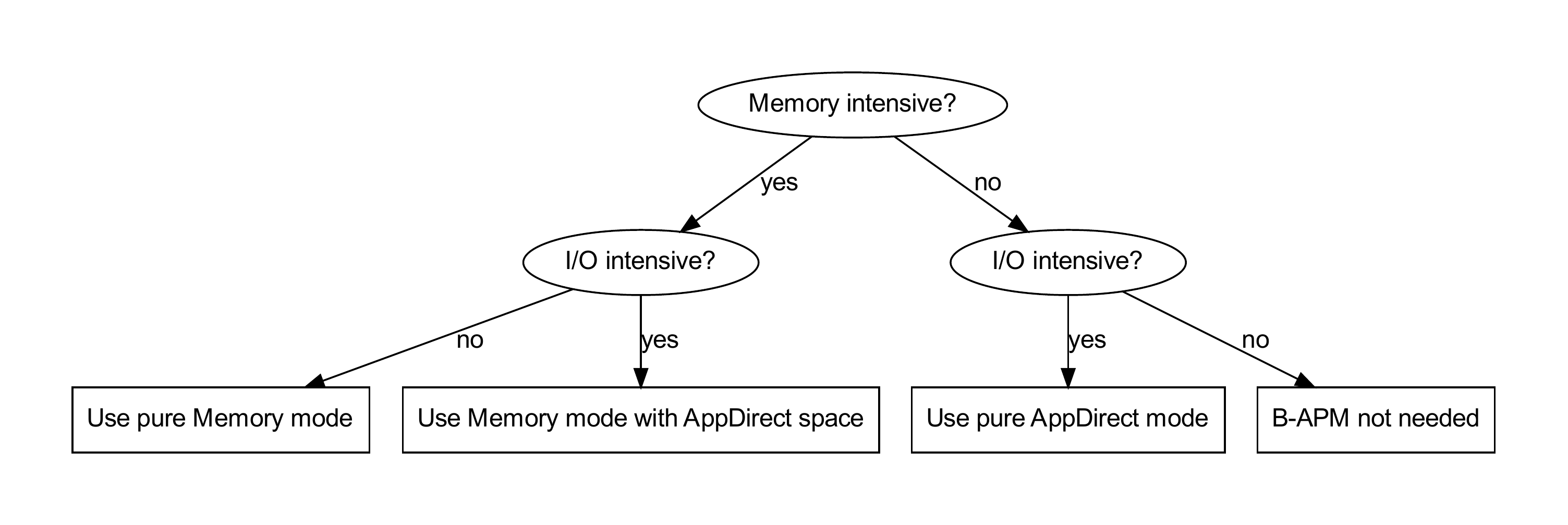}
\caption{Decision tree to choose the configuration of the B-APM based on the needs of an application}
\label{stream}
\end{figure*}
\baselineskip=18pt plus.2pt minus.2pt
\parskip=0pt plus.2pt minus0.2pt

\setcounter{figure}{9}
\begin{figure*}[!htb]
\centering
\includegraphics[scale=0.45]{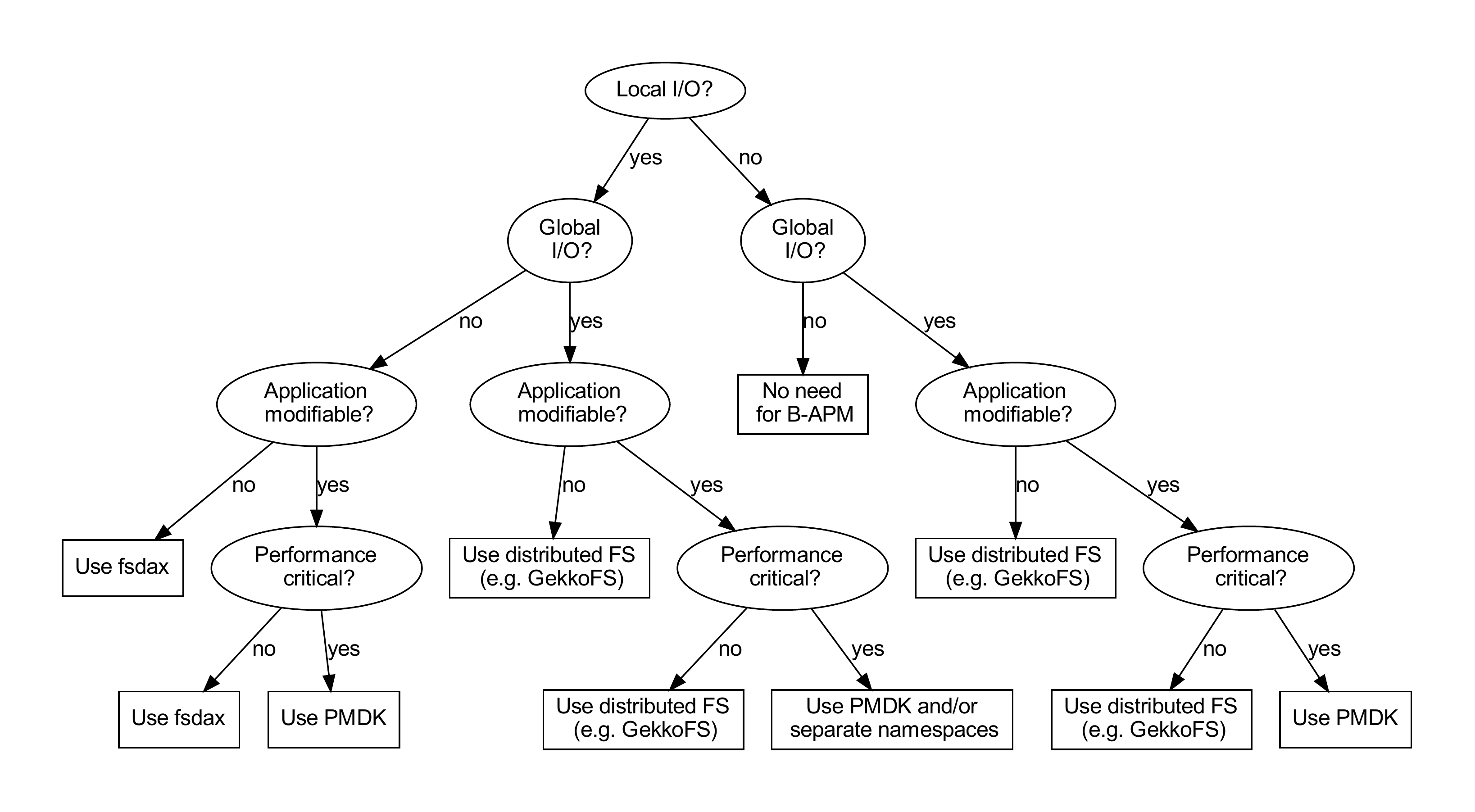}
\caption{Options for using on the AppDirect space for I/O intensive applications, depending on the I/O patterns, whether the application can be modified, and whether optimal performance is desired}
\label{stream}
\end{figure*}
\baselineskip=18pt plus.2pt minus.2pt
\parskip=0pt plus.2pt minus0.2pt

\section{Using the Decision Trees}
This section demonstrates how the decision trees introduced in Section 4.4 might be used to select appropriate B-APM configurations given an application's resource needs. Note that these are examples given for the purpose of illustration and they are not necessarily the only possible, or indeed best, solutions. What constitutes ``appropriate'' or ``optimal'' for an application might vary and depends on what criterion it is being optimised for. Possible optimisation targets could be shortest time to solution, high parallel efficiency, minimal power or minimal energy.  

\subsection{Memory-intensive meshing}
The snappyHexMesh utility, part of the OpenFOAM framework referred to in Section~\ref{sec:locstor}, is a meshing application that refines a decomposed coarse mesh to more accurately represent a given geometry. It is written in C++ and parallelised using MPI. For large geometries, snappyHexMesh (in line with many meshing applications) is memory intensive, but at the same time scalability can suffer from poor load-balancing, in particular for irregular geometries. In this example, snappyHexMesh is used to refine a complex mesh with 105 million cells. On the NEXTGenIO system, without using any of the available B-APM, a single node (48 cores and 192GB RAM) meshing attempt fails with an ``out of memory'' error. Using two nodes (96 cores and 384GB RAM) the application runs to completion, taking 6,566s. Note that the performance variability between different runs is in the order of a few tens of seconds, and we reported the fastest achieved time here. A performance profile of the application shows that more than 60\% of the total runtime on 96 cores is spent in the MPI library, with I/O being negligible at less than 1\%. The high proportion of time spent in MPI points towards poor load balancing and thus being able to reduce the number of cores for the application should improve efficiency.

The decision tree in Figure 9 tells us that for an application that is memory intensive, but \emph{not} I/O intensive, using a pure Memory mode setup might be the correct solution. Here the desire is to be able to run the application on a smaller number of nodes with an aim to improving the parallel efficiency. As the use of Memory mode is entirely transparent to the user, no other changes need to be made. Rerunning snappyHexMesh on a compute node that has been configured to run in Memory mode (i.e. using 3TB main memory and 192GB of DRAM cache) yields a runtime of 11,637s. This is considerably slower in terms of absolute wall clock time, and if the goal is to perform the meshing in the shortest time possible, this would not be the optimal configuration. However, in terms of efficiency (i.e. looking at the runtime relative to the resources being used) using Memory mode represents a 15\% improvement for this case. If the goal is to complete as many meshing jobs as possible given a fixed amount of resources, Memory mode presents a benefit due to the poor scaling behaviour of the application. Even on a single node the amount of time spent in the MPI library remains high at just under 50\%, which suggests that further load balancing optimisations would be advised to improve the performance.

\subsection{Ensemble of single node runs}
The Met Office NERC Cloud model (MONC) is a high-resolution modelling framework that uses large eddy simulations to study the physics of turbulent flows~\cite{monc}. In particular, it is used to compute parameters for numerical weather prediction models. The code is written in Fortran 2008 and parallelised using MPI. An interesting aspect of this application is that it has two sets of processes running simultaneously: compute processes, which run the main simulation, and I/O server processes. The latter are processes that are dedicated to performing the I/O, either writing diagnostics or checkpoint files. The I/O server processes jointly write the output files in parallel. MONC is inherently designed to cope well with writing large amounts of data and to allow the simulation to progress as much as possible even when there is contention on the I/O. However, this does not mean that the application is immune to I/O lag - one such scenario can be observed where a number of single-node instances of MONC are run at the same time. Here, we run 10 simulations concurrently, and each simulation is set up to complete when 100s of ``model'' time has been reached. Depending on the size of the simulation and the timestep delta, each second of model time can take a significant amount of time to compute both in terms of wall clock time and number of timesteps. Every 100 timesteps, a checkpoint file is written - the size of the checkpoint files grows with the simulation time and in our experiment, the file size starts at 500MB and grows to 4.5GB, with a total of 16 files being written. The mean wall clock time for each of the 10 simulations to reach 100 model seconds is 626s, with a minimum time of 547s for the fastest and 710s for the slowest. Even though all the simulations perform that same computation, the variability in runtime is significant (approximately 13\% for this example).

\setcounter{table}{2}
\tabcolsep 9pt
\renewcommand\arraystretch{1.3}
\begin{table*}[!htb]
\centering
\caption{\label{3} Summary of the applications and related usage scenarios showcased in this paper. For each application, the configuration choices are influenced by the specific optimisation targets for the application. Note that the choices are specific to the test cases and experiments shown here, and might differ for alternative setups}\vspace{-2mm}
{\footnotesize
\begin{tabular*}{\linewidth}{l|lllll|l}
\toprule
  \multirow{2}{*}{Application} & Memory & I/O & I/O access & Code changes & I/O perf. & Chosen \\
 & intensive? & intensive? & pattern? & possible? & critical? & configuration \\ 
\midrule
\midrule
CASTEP & yes & no & n/a & n/a & n/a & Memory mode \\
\hline
snappyHexMesh & yes & no & n/a & n/a & n/a & Memory mode \\
\hline
\multirow{2}{*}{simpleFoam} & \multirow{2}{*}{no} & \multirow{2}{*}{yes} & \multirow{2}{*}{local} & yes in principle, & \multirow{2}{*}{yes} & \multirow{2}{*}{AppDirect (fdsdax)} \\
 &  &  & & but undesirable & &  \\
\hline
IO500 & no & yes & local + global & no & yes & AppDirect (GekkoFS) \\
\hline
MONC ensemble & no & yes & local & yes & no & AppDirect (fsdax) \\
\hline
FDB5 & no & yes & global & yes & yes & AppDirect (PMDK) \\
 \bottomrule
\end{tabular*}
}
\end{table*}

If this performance variability poses a problem, using faster I/O might be a solution. For this particular set of simulations, memory consumption is not a concern as all the individual simulations are small. Looking at Figure 9, it is therefore clear that a pure AppDirect configuration should be chosen. Next, Figure 10 guides the decision for how B-APM might be used. In the case of MONC, the I/O server processes perform global I/O, however in this particular setup (with single node simulations) that global I/O becomes in fact local. The application is modifiable, however here we are not so much interested in achieving the ultimate performance rather than reducing the variability of the runtimes across the individual simulations. Following the decision tree, we therefore choose to simply write the output data to fsdax and copy it back to Lustre at the end of the simulation.

We then run the same 10 simulations as before, but with the new output target, fsdax. The average runtime across ten simulations is now 550s (down from 626s), with a minimum time of 517s for the fastest and a maximum tine of 568s for the slowest simulation. There is still performance variability, however this has been reduced from 13\% to around 5\%. The fastest runtime when using Lustre as an output target is in line with the average for B-APM: 547s for Lustre vs 550s for fsdax, which shows that the I/O server can in principle achieve the same performance with both Lustre and fsdax. However the big reduction in performance variability comes from that fact that writing directly to B-APM is a purely node-local operation that does not involve the network, which is a shared resource and thus can be a source of contention.

\section{Importance of persistence}
One of the key characteristics of B-APM, namely the fact that the technology is persistent, has not been mentioned as being vital in any of the usage scenarios outlined above, because (in truth) it is not important for most HPC or HPDA applications which run on system that have a network attached filesystem that can keep a copy of any data. However, there are couple of scenarios where persistence might be beneficial:

$\bullet$ Automatic snapshots of intermediary results: traditionally, check-pointing is used to periodically save data to disk. Using the persistence of B-APM, it is possible to keep continuous snapshots of the most recent results without impacting performance, ensuring that even if a simulation is killed because of a loss of power, the data is kept safe and the simulation can resume seamlessly.

$\bullet$ Large datasets for long term use: rather than keeping large datasets on network attached storage, it is possible to place them directly onto the B-APM on compute nodes and keep them there long term, potentially to be accessed by multiple users. This is of course already possible with node-local storage such as SSDs, however the performance benefits of B-APM make this an even more attractive proposition. 

\section{Conclusions}
B-APM has the potential to transform the memory and storage space. With byte-addressability, and with performance close to DRAM, B-APM can provide solutions to a range of HPC usage scenarios with high memory and I/O performance demands. Table 3 summarises the scenarios presented in this paper. B-APM, and DCPMM in particular, allows a hardware-homogeneous system to be configured in a highly flexible (heterogeneous) way so that the needs of memory and I/O intensive applications can be supported individually using the same hardware configured for different application needs. This paper only shows some of the most common usage scenarios in the HPC and data analytics space, but the potential of the technology is much greater. As more B-APM products come on the market, more usage scenarios will be explored and supported by system software, which means that more innovative ways of supporting application and workflow needs will open up, benefiting memory and I/O intensive applications across the board.

\vspace{5mm}

\noindent\parbox{8.3cm}{\parpic{\includegraphics[width=2.5cm]{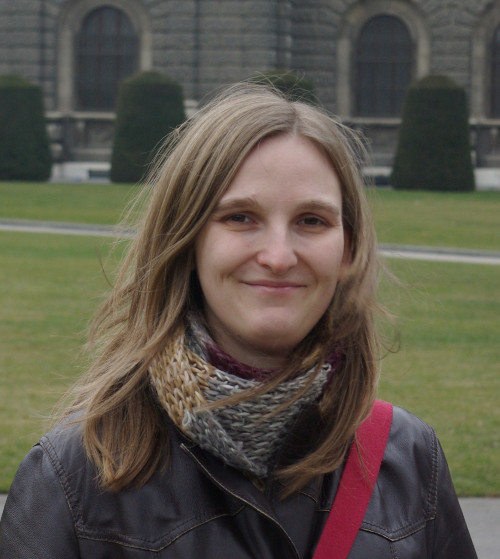}}{\small\quad{\bf Mich\`{e}le Weiland} is a Senior Research Fellow at EPCC, the supercomputing centre at the University of Edinburgh. She specialises in novel technologies for extreme scale parallel computing, leading EPCC's technical work in the ASiMoV Strategic Prosperity
Partnership with Rolls-Royce. She was responsible for managing the technical work as part of the EU H2020 project NEXTGenIO. She is the EPCC Principal Investigator on a number of research grants, including the EC Horizon 2020 projects HPC-WE and SAGE2.}\\[1mm]}

\noindent\parbox{8.3cm}{\parpic{\includegraphics [width=2.5cm]{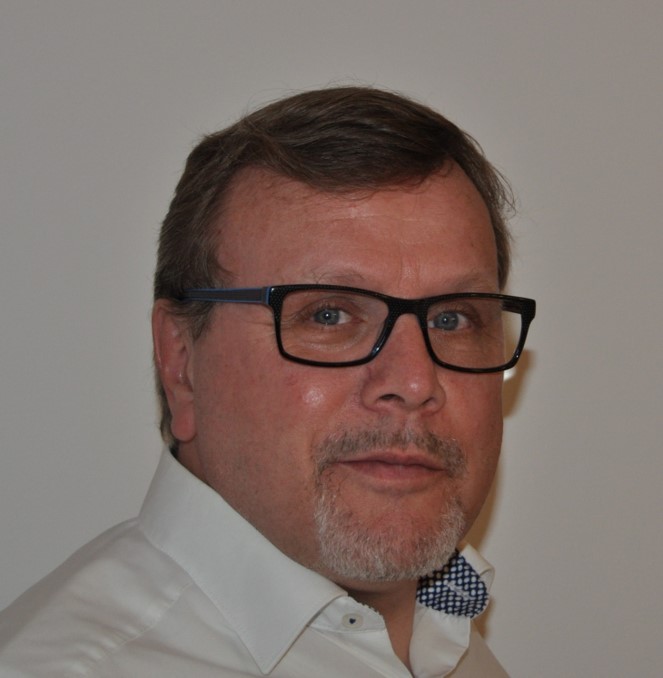}}{\small\quad {\bf Bernhard Hom\"{o}lle}  is the head of the HPC competence center at SVA System Vertrieb Alexander GmbH in Germany. He has a long history (more than 30 years from Nixdorf Computers to Fujitsu) in the development of server systems and was responsible Principal Architect for Fujitsu PRIMGERY Sever and Distinguished Engineer at Fujitsu. For the EU H2020 project NEXTGenIO led the work package ``Prototype Hardware Development''.   }\\[1mm]}

\label{last-page}
\end{multicols}
\label{last-page}

\begin{thebibliography}{99}
\footnotesize
\itemsep=-3pt plus.2pt minus.2pt
\baselineskip=13pt plus.2pt minus.2pt

\bibitem{jackson} A. Jackson, M. Weiland, M. Parsons, and B. Hom\"olle, ``An Architecture for High Performance Computing and Data Systems Using Byte-Addressable Persistent Memory'', in High Performance Computing, vol. 11887, M. Weiland, G. Juckeland, S. Alam, and H. Jagode, Eds. Cham: Springer International Publishing, 2019, pp. 258--274

\bibitem{izraelevitz} J. Izraelevitz et al., ``Basic Performance Measurements of the Intel Optane {DC} Persistent Memory Module'', arXiv, http://arxiv.org/abs/1903.05714, March 2019

\bibitem{patil} O. Patil et al., ``Performance Characterization of a DRAM-NVM Hybrid Memory Architecture for HPC Applications Using Intel Optane DC Persistent Memory Modules'', in Proceedings of the International Symposium on Memory Systems (MEMSYS'19). Association for Computing Machinery, 2019, pp.288--303

\bibitem{mason} T. Mason, T. D. Doudali, M. Seltzer and A. Gavrilovska, ``Unexpected Performance of Intel Optane DC Persistent Memory'', in IEEE Computer Architecture Letters, vol. 19, no. 1, pp. 55-58, 1 Jan--June 2020

\bibitem{castep} S.J. Clark et al.,``First principles methods using {CASTEP}'', Z. Kristall, 2005, vol. 220, pp.567--570

\bibitem{weilandSC} M. Weiland et al., ``An early evaluation of Intel's Optane DC persistent memory module and its impact on high-performance scientific applications'', in Proceedings of the International Conference for High Performance Computing, Networking, Storage and Analysis, Denver Colorado, Nov. 2019, pp. 1--19, doi: 10.1145/3295500.3356159

\bibitem{vef} M.-A. Vef et al., ``GekkoFS -- A Temporary Burst Buffer File System for HPC Applications'', J. Comput. Sci. Technol., vol. 35, no. 1, pp. 72--91, Jan. 2020

\bibitem{brinkmann} A. Brinkmann et al., ``Ad Hoc File Systems for High-Performance Computing'', J. Comput. Sci. Technol., vol. 35, no. 1, pp. 4--26, Jan. 2020

\bibitem{fdb2017} S. Smart, T. Quintino and B. Raoult, ``A Scalable Object Store for Meteorological and Climate Data'', in Proceedings of the Platform for Advanced Scientific Computing Conference (PASC'17), 2017.

\bibitem{fdb2019} S. Smart, T. Quintino and B. Raoult, ``A High-Performance Distributed Object-Store for Exascale Numerical Weather Prediction and Climate'', in Proceedings of the Platform for Advanced Scientific Computing Conference (PASC'19), 2019

\bibitem{weilandJSFI} M. Weiland, A. Jackson, N. Johnson and M.Parsons, ``Exploiting the Performance Benefits of Storage Class Memory for HPC and HPDA Workflows'', Journal of Supercomputing Frontiers and Innovations, vol. 5, no. 1, Mar. 2018, doi: 10.14529/jsfi180105

\bibitem{norns} A. Miranda, A. Jackson, T. Tocci, I. Panourgias, and R. Nou, ``NORNS: Extending Slurm to Support Data-Driven Workflows through Asynchronous Data Staging'', in 2019 IEEE International Conference on Cluster Computing (CLUSTER), Albuquerque, NM, USA, Sep. 2019, pp. 1--12

\bibitem{monc} N. Brown et al. ``A highly scalable Met Office NERC Cloud model'', Proceedings of the 3rd International Conference on Exascale Applications and Software, Edinburgh, July 2015

\end{thebibliography}
\end{document}